\documentclass[aps,pra,twocolumn,superscriptaddress,floatfix,showpacs]{revtex4-1}

\usepackage{graphicx}
\usepackage{amssymb}
\usepackage{color}
\usepackage{psfrag}
\usepackage{ifsym}
\usepackage{hyperref}

\definecolor{pink}{rgb}{1,0.078,0.57}

\newcommand{\ket}[2] {| #1 \rangle_{#2}}
\newcommand{\bra}[2] {\langle #1 |_{#2}}

\begin{document}

\title{Shaping the joint spectrum of down-converted photons through optimized custom poling}

\author{Annamaria Dosseva}
\affiliation{Institute for Quantum Computing, University of Waterloo, Waterloo, Ontario N2L 3G1, Canada}
\author{\L{}ukasz Cincio}
\affiliation{Perimeter Institute for Theoretical Physics, Waterloo, Ontario, N2L 2Y5, Canada}
\author{Agata M. Bra\'nczyk}
\email{abranczyk@perimeterinstitute.ca} 
\affiliation{Perimeter Institute for Theoretical Physics, Waterloo, Ontario, N2L 2Y5, Canada}

\begin{abstract}
We present a scheme for engineering the joint spectrum of photon pairs created via spontaneous parametric down conversion. Our method relies on customizing the poling configuration of a quasi-phase-matched crystal.  We use simulated annealing to find an optimized poling configuration which allows almost arbitrary shaping of the crystal's phase-matching function. This has direct application in the creation of pure single photons---currently one of the most important goals of single-photon quantum optics. We describe the general algorithm and provide code, written in \verb+C+++, that outputs an optimized poling configuration given specific experimental parameters. 
\end{abstract}

\maketitle

\section{Introduction}\label{intro}

The generation of pure nonclassical states of light is one of the most important goals of optical quantum information science \cite{Kok2010}. A popular and versatile source of nonclassical light is spontaneous parametric down-conversion (SPDC)  \cite{Christ2013}---a nonlinear process that converts  high-energy photons into pairs of lower energy photons. SPDC has been employed in the generation of squeezed light \cite{Wu1986}, Schr\"odinger kitten states \cite{Ourjoumtsev2006} and entangled photons \cite{Kwiat1999a}, and is  the most widely used technique for generating single photons \cite{Christ2013}. 

Sources based on SPDC have widespread application in quantum computation \cite{Kok2007}, quantum communication \cite{Gisin2007}, and quantum metrology \cite{Higgins2007,Nagata2007}, as well as in more specialized areas such as quantum imaging \cite{Brida2010}, quantum lithography \cite{Boto2000}, or optical coherence tomography \cite{Nasr2008}. The ability to control the characteristics of quantum states of light becomes increasingly important as these applications mature.

In general, photon pairs generated via SPDC are correlated in frequency and are described by a \emph{joint spectral amplitude} (JSA), which is determined by the properties of the incident pump field and the material properties of the nonlinear crystal used to mediate the down-conversion process.  In this paper, we focus on controlling the \emph{spectral} properties of down-conversion sources. 

One particularly challenging but important application of spectral shaping is in the production of single photons, which are generated from SPDC photon-pair sources through a heralding process, whereby the detection of one photon heralds the presence of the other.  Spectral correlations between the photons degrade the spectral purity of heralded photons and are therefore undesirable. 

The simplest method for modifying the JSA's shape  is filtering, which can reduce spectral correlations. But because it introduces a spectrally dependent loss which acts on the individual photons independently, spectral filtering can degrade the quantum state's photon-number purity \cite{Branczyk2010,Christ2014}.

More sophisticated methods involve shaping the spectrum at the source using techniques such as quasi-phasematching (QPM) \cite{Fejer1992,Imeshev2001}. Because such methods act on both photons in a pair simultaneously, they do not affect the quantum state's photon-number purity. QPM can be achieved through a technique known as periodic poling, where the nonlinear medium is constructed from individual domains of birefringent material with alternating orientation, see Fig. 1. Chirped gratings have been employed for pulse compression in second-harmonic generation, as well as the generation of ultrabroad-spectrum, top-hat shaped photons for optical coherence tomography \cite{Nasr2008}. Correlations in the JSA can be reduced by using periodic poling in conjunction with group-velocity matching \cite{Grice2001,Raymer2005,URen2005,Garay-Palmett2007,Mosley2008,Smith2009,Vicent2010,Gerrits2011}, however, the extent to which spectral separability can be achieved is limited by the crystal's inherent sinc-type phasematching function, which manifests itself in the JSA as undesirable diagonal side-lobes, see Fig 2 a).

\begin{figure}[t]
\begin{center}
\includegraphics[width=\columnwidth]{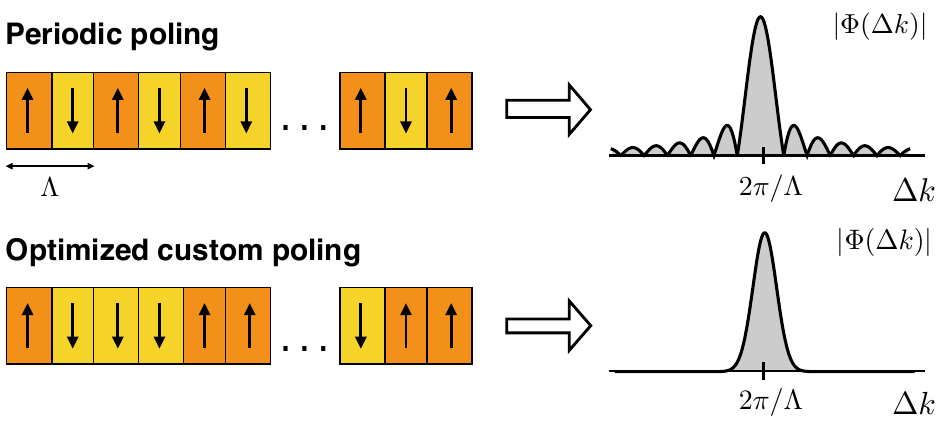}
\caption{(Color online) Periodic poling produces a sinc-shaped phase-matching function. Alternative phase-matching functions, e.g., Gaussian, can be  generated by customizing the poling configuration. The phase mismatch, $\Delta k$, has units of rad/m. }
\end{center}
\label{fig:schematic}
\end{figure}

A technique for shaping the phasematching function using nontrivial QPM was proposed by  Bra\'nczyk \emph{et al.} \cite{Branczyk2011}, who showed that modulation of the nonlinearity profile of a down-conversion crystal can drastically reduce side lobes in the JSA. In this method, a discretized approximation to the desired nonlinearity profile was achieved using higher-order poling. Dixon \emph{et al.} \cite{Dixon2013} proposed an alternative method for spectral decorrelation, in which the crystal's duty-cycle pattern is customized while the grating period is fixed. 

In this paper, we introduce a method in which we directly manipulate the domain orientations while keeping their widths fixed (Fig. 1, bottom). In contrast with the methods proposed in \cite{Branczyk2011,Dixon2013}, our method allows almost arbitrary shaping of the phase-matching function, which provides flexibility in designing the maximum nonlinearity for a given phase-matching function width. Furthermore, in contrast with the method  in \cite{Dixon2013}, we fix the width of each domain in the grating, which preserves the phase-matching properties of the crystal---a stringent requirement in many experiments. 

While we focus on correlations in the spectral domain, spatial correlations also exist and similar concepts have been discussed in \cite{Kolobov1999, Law2004, Gatti2009,Walborn2010}.

\section{Joint spectral amplitude of downconverted pairs}

The SPDC process mediates the conversion of high-energy pump photons in mode $p$ into pairs of lower energy photons in modes $a$ and $b$. This process satisfies energy and momentum conservation according to: $\omega_p=\omega_a+\omega_b$ and $k_p(\omega_p)=k_a(\omega_a)+k_b(\omega_b)$, where $\omega_j$ is the frequency in mode $j$, and $k_{j}(\omega)=n_{j}(\omega)\omega/c$ is the wave vector associated with the polarization of mode $j$, evaluated at frequency $\omega$.

Theoretically, the two-photon state generated via SPDC can be described by \cite{Grice1997}
\begin{equation}\label{eq:twophot}
\ket{\psi}{}=\int \mathrm{d}\omega_{a} \int \mathrm{d}\omega_{b}f(\omega_{a},\omega_{b})\ket{\omega_{a}}{a}\ket{\omega_{b}}{b}\,,
\end{equation}
where $\ket{\omega_{i}}{j}$ is a one-photon Fock state of frequency $\omega_{i}$ prepared in mode $j$. The JSA, $f(\omega_{a},\omega_{b})=\alpha(\omega_{p})\Phi(\omega_a,\omega_b)$, characterizes the joint spectrum of the two photons.  The spectral properties of downconverted photons can be manipulated via the pump beam spectral amplitude function $\alpha(\omega_{p})$ and the phase-matching function $\Phi(\omega_{a},\omega_{b})$.  

The JSA \emph{cannot}, in general, be factorized into a product of separable single-photon spectral amplitudes, ${u}(\omega_{a}){v}(\omega_{b})$, however, it can always be decomposed into a weighted sum of separable single-photon spectral amplitudes, $f(\omega_{a},\omega_{b})=\sum_{k}b_k{u}_k(\omega_{a}){v}_k(\omega_{b})$, known as the Schmidt decomposition. The functions ${u}_k(\omega_{a})$ and ${v}_k(\omega_{b})$ each form a discrete basis of complex orthonormal functions and the Schmidt coefficients ${b}_{k}$ are real and satisfy $\sum_{k}{b}_{k}^2=1$ if $f(\omega_{a},\omega_{b})$ is normalized. 

In terms of the Schmidt decomposition, the downconverted state can be written as
\begin{equation}\label{eq:schmidt}
\ket{\psi}{}=\sum_{k}{}b_k\ket{u_k}{a}\ket{v_k}{b}\,,
\end{equation}
where $\ket{u_k}{}=\int \mathrm{d}\omega_a u_k(\omega_a)\ket{\omega_a}{a}$ and $\ket{v_k}{}=\int \mathrm{d}\omega_b v_k(\omega_b)\ket{\omega_b}{b}$.

The degree of correlation of a pure bipartite state, i.e., entanglement, can then be characterized by the entropy of entanglement \cite{Bennett1996}. In terms of the Schmidt coefficients, this is given as $E=-\sum_{k}{b}_{k}^2\log_2({b}_{k}^2)$. For a completely decorrelated JSA, the Schmidt decomposition contains only one term, i.e., $b_1=1$, and the entropy of entanglement is $E=0$.

\section{Purity of heralded single photons}\label{sec:pur}

Consider any pure entangled bipartite state. The reduced density matrix of either subsystem will necessarily be mixed. The archetypical example of this is the Bell state---a Bell state is maximally entangled, and thus, the subsystems are maximally mixed. 

Similarly, spectral correlations in the bi-photon spectral amplitude will necessarily reduce the spectral purity of the individual photons. Consider the state in  Eq (\ref{eq:schmidt}). Given detection of a single photon in mode $b$ by a detector that does not provide any spectral information, the single-photon state in mode $a$ can be written as $\rho_{a}=\sum_{k}{b}_{k}^2\ket{{u}_{k}}{a}\bra{{u}_{k}}{a}$. The purity of the reduced density matrix is given by $P=\mathrm{Tr}[\rho_{a}^2]=\sum_{k}{b}_{k}^4$. If the Schmidt decomposition has only one non-zero Schmidt coefficient, $\rho_{a}$ is a pure state and $P=1$. To increase the purity of the heralded photon, one should therefore aim to reduce correlations such that the Schmidt decomposition has only one non-zero Schmidt coefficient. 

Group-velocity-matching reduces JSA correlations by orienting the functions $\Phi(\omega_{a},\omega_{b})$ and $\alpha(\omega_a+\omega_b)$ perpendicular to each other and matching their widths as closely as possible. This occurs when the group velocities satisfy $k'_p=(k'_a+k'_b)/2$, where $k'_j=\partial k_j(\omega)/\partial \omega|_{\omega=\bar{\omega}_j}$ and $\bar{\omega}_j$ are the central frequencies \cite{Grice2001,Raymer2005,URen2005,Garay-Palmett2007,Mosley2008,Smith2009,Vicent2010,Gerrits2011}.  Experiments in this regime  typically employ a  periodically-poled crystal which leads to a sinc-type phase-matching function that manifests itself in the JSA as undesirable diagonal side-lobes. In this paper, we focus on directly shaping the phase-matching function to remove the side lobes, thus increasing the separability of the joint spectral amplitude. 

\section{Shaping the phase-matching function}

The phase-matching function is related to the nonlinearity profile of the crystal via the Fourier transform \cite{Grice1997}
\begin{equation}\label{eq:phase-matching function}
\Phi(\omega_{a},\omega_{b})\propto\frac{1}{L}\int_{-\infty}^{\infty}\chi(z)e^{-i\Delta k(\omega_{a},\omega_{b})z}\mathrm{d}z\,,
\end{equation} where $L$ is the crystal length, $\chi(z)$ represents the nonlinear optical coupling, and where phase matching within the crystal can be described by the phase-mismatch $\Delta k(\omega_{a},\omega_{b})=k_{p}(\omega_{a}+\omega_{b})-k_{a}(\omega_{a})-k_{b}(\omega_{b})$. We will use $\Delta k$ as shorthand for $\Delta k(\omega_{a},\omega_{b})$.

In principle, an arbitrary $\Phi(\omega_{a},\omega_{b})$ could be realized with appropriate design of the crystal's nonlinearity profile such that it corresponds to the Fourier transform of the desired phase-matching function. Unfortunately, it is non-trivial to directly change the material properties of a nonlinear crystal, and different methods must be used.

Consider a nonlinear medium composed of $N$ domains of birefringent material of length $l_c$, where each domain can be oriented either \emph{up} or \emph{down}. A flip in the orientation of the domain introduces a phase shift of $\pi$. The nonlinearity profile $\chi(z)$ in such a crystal  is a discontinuous function that only takes on values of $\pm\chi_0$. The phase-matching function for the entire crystal is a linear superposition of phase-matching functions for individual crystal domains:
\begin{eqnarray}
\Phi(\omega_{a},\omega_{b})&\propto&\frac{\chi_{0}}{L}\sum_{{n}=1}^{N} s_{n}\int_{-\infty}^{\infty}\mathrm{rect}\left(\frac{{z}-{z}_{n}}{{l}_{c}}\right)e^{-i\Delta kz}\mathrm{d}z~~\\\label{eq:sum}
&\propto&\frac{\chi_{0}{l}_{c}}{L}\mathrm{sinc}   \left(\frac{\Delta k {l}_{c}}{2}\right)\sum_{{n}=1}^{N} s_{n} e^{-i \Delta k z_{n}}\,,
\end{eqnarray}
 where ${s}_{n}$ accounts for the phase shift due to the orientation of the domain and ${z}_{n}=({n}-\frac{1}{2}){l}_{c}$ specifies the origin of the $n$th domain. The special case, where ${s}_{n}=e^{i n \pi}=(-1)^n$, corresponds to a periodically poled crystal. In this case, constructive interference near the point $\Delta k=2\pi/\Lambda$, where  $\Lambda=2{l}_{c}$ is the poling period,  produces a phase-matching function that approximates
\begin{eqnarray}\label{eq:phiapp}
\Phi(\omega_{a},\omega_{b})\propto\mathrm{sinc}\left(\frac{\left(\Delta k-\frac{2\pi}{\Lambda}\right) L}{2}\right)\,.
\end{eqnarray}

Since, in principle, the domains can take on other configurations, a natural question to consider is whether the phase-matching function can be tailored by manipulating the relative orientations of the individual domains in a non-trivial way. This was demonstrated by Bra\'nczyk \emph{et al.} \cite{Branczyk2011}, where the authors designed and experimentally verified a Gaussian phase-matching function for a 1 cm potassium titanyl phosphate (KTP) crystal with \mbox{$\Lambda=10.85\mu$m}.  But the design in \cite{Branczyk2011} was specific to a particular set of parameters, and cannot easily be generalized. 

In this paper, we introduce a \emph{general} technique for optimizing the domain orientations in order to achieve a desired phase-matching function. Since periodically poled crystals are optimized for source brightness (around a specific centre frequency),  a custom poled crystal will necessarily generate fewer pairs compared to a periodically poled crystal of the same length pumped with the same pump.

In the next section, we describe the algorithm used for this optimization.

\section{The algorithm} \label{sec:alg}

The task of finding an optimal domain configuration is formulated in terms of discrete optimization. The variables are the $N$ possible domain orientations $s_n$, constrained as $s_n \in \{-1,1\}$ and the solution space consists of $2^N$ possible crystal configurations. Neighbour configurations are defined as those which differ by exactly one domain orientation. Since there are $N$ domains which could be flipped, any particular configuration has $N$ neighbours.

Each domain configuration ${\bf s} \equiv (s_1,\ldots,s_N)$  has a corresponding nonlinearity profile, with a phase-matching function $\Phi_{\bf s}(\Delta k)$ given by the Fourier transform of this nonlinearity profile. The task is to find the crystal alignment ${\bf s}_0$ which yields a $\Phi_{{\bf s}_0}(\Delta k)$ closest to some target function, $\Phi_{\mathrm{target}}(\Delta k)$, on a specified range $[a,b]$.

We define a cost function
\begin{equation}
d_{\bf s}= \sum _{m=1}^{M} | \Phi_{\bf s}(\Delta k_m)- \Phi_{\mathrm{target}}(\Delta k_m)  |
\end{equation}
as a measure of the distance between $\Phi_{\bf s}(\Delta k)$ and $\Phi_{\mathrm{target}}(\Delta k)$. This distance is measured by selecting sufficiently many, say $M=2000$, points  $\Delta k_m$ in the range $[a,b]$. The objective is to minimize $d_{\bf s}$.

Because  $N$ will typically range from approximately one hundred to a few thousand, the solution space is  too large to check each possible configuration with currently available computing resources. It is also not convex, meaning that there exist configurations which are locally optimal---that is, each neighbour is a worse configuration---but not globally optimal; one can not simply move from neighbour to neighbour, always selecting the better configuration. We solve this problem by using simulated annealing \cite{Van-Laarhoven1987}, a method which accepts worse configurations with a decaying probability $p$, and better configurations with some probability $q$ that is close to $1$.

Starting with either a random or preselected domain configuration, the algorithm iterates through randomly-chosen neighbour configurations, deciding whether to flip a given domain or not. If ${\bf s}$ is the current configuration, and ${\bf s}'$ is a neighbour configuration being considered, then when $d_{{\bf s}'}\geq d_{\bf s}$, the algorithm moves to (worse) ${\bf s}'$ with probability $p_i$ at iteration $i$; however, when $ d_{{\bf s}'}< d_{\bf s}$, the algorithm moves to (better) ${\bf s}'$ with probability $q$. 

The probability of accepting a worse configuration at iteration $i$ is given by $p_i = (h_i / A)\times(d_{\bf s}/ d_{{\bf s}'})$, where $h_i$ is a decaying function known as the `heat function' (in analogy with physical annealing), and $A$ is roughly the number of domains explored before the algorithm would hit a local minimum if $p_i$ were set to zero. The heat function is chosen to be $h_i = 2\times2^{-i/J}-1$, where $J$ is the total number of iterations.  

Finally, we use $q=0.999$, determined heuristically. Any number close, but not equal to $1$ tends to give good results. Parameters $q$ and $A$ can be adjusted further to improve performance of the algorithm.

After $J$ iterations, the algorithm performs an additional optimization by systematically sweeping through the crystal (say, from left to right), flipping each domain and only keeping the new configuration if $ d_{{\bf s}'}< d_{\bf s}$. The algorithm stops when an entire sweep does not reduce the cost function.

The algorithm is probabilistic, and may therefore occasionally yield an unsatisfactory solution. This can be ruled out by running the algorithm several times and comparing the results. 

The solutions are not unique, in that different configurations can yield the same value of the cost function.

\subsection{Choosing appropriate input parameters}\label{sec:app}

Here, we provide some rules of thumb for selecting input parameters. We imagine a situation where the experimenter knows the desired target function $\Phi_{\mathrm{target}}(\Delta k)$ and wants to choose the crystal parameters $l_c$ and $L$ appropriately for best performance of the algorithm. If these parameters are chosen arbitrarily, the algorithm will still find an optimized solution, however, this solution may not be sufficiently close to the desired function. 

A good value for the parameter $l_c$ can be determined by identifying the peak of the desired phase-matching function, let's call it $\Delta k_{peak}$, and setting $l_c=\pi/\Delta k_{peak}$ (if there is no identifiable peak,  $\Delta k_{\mathrm{peak}}$ can be chosen to be a point in the range $[a,b]$). The parameter $N$ should be chosen large enough, such that the Fourier transform of $\Phi_{\mathrm{target}}(\Delta k)$ has strong support in the range $[-Nl_c/2,Nl_c/2]$, but not much larger. When inputting the phase-matching function into the algorithm, the height of $\Phi_{\mathrm{target}}(\Delta k)$ should be between 0 and $2Nl_c/\pi$. The  ideal height---that is, the height for which the algorithm performs best---seems to have a non-trivial dependence on $N$ and $l_c$, as well as the form of the target function. However, the dependence isn't overly sensitive: a deviation of $5\%$ from the ideal height still works very well. It is relatively simple to identify the ideal height through trial and error.

\begin{figure*}[t]
\label{fig:2}
\begin{center}
\includegraphics[width=\textwidth]{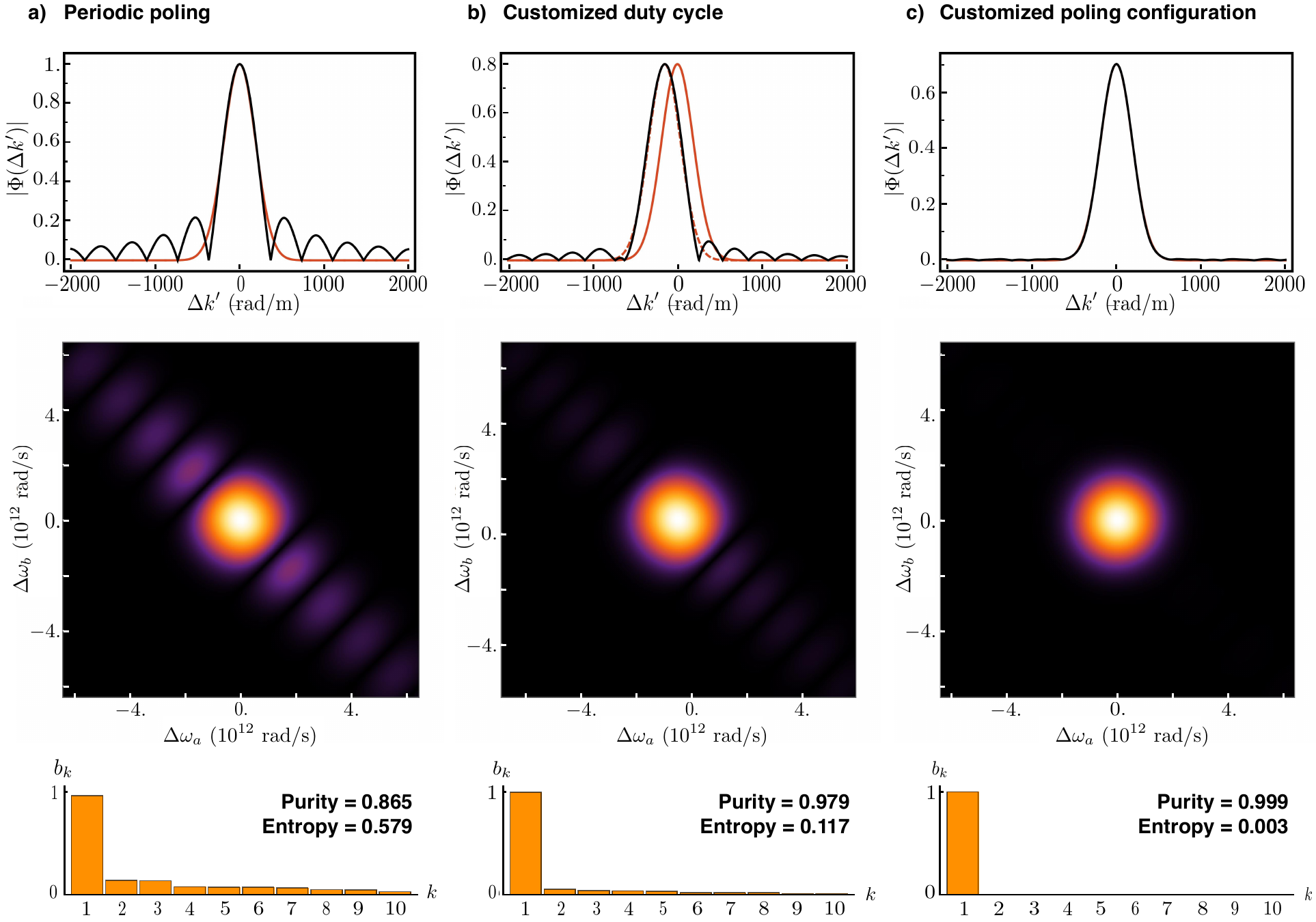}
\caption{(Color online) Comparison of phase-matching functions (top), joint spectral amplitudes (middle) and Schmidt coefficients (bottom), which are all dimensionless, for different apodization schemes. The  Gaussian target function is also shown in red. a) Standard first-order periodically poled crystal (no apodization). We define the peak nonlinearity of this crystal to be unity and scale all other peak nonlinearities accordingly. b) Customized duty cycle method proposed by Dixon \emph{et al.} \cite{Dixon2013}. In this scheme, the peak center is shifted, but this does not affect the photon-purity. The dashed, red line shows the Gaussian target function shifted for comparison. c) Custom-poled crystal generated using simulated annealing. We used a randomized initial trial configuration, then ran the algorithm with $2\times10^5$ iterations, using $q=0.999$ and $A=1000$. We used \mbox{$M=2001$}  samples to compute the objective function $d_{\bf s}$, over the range \mbox{$[a,b]=[2\pi(1-0.025)/\Lambda, 2\pi(1+0.025)/\Lambda]$}. The height of the target function was $H=0.8Nl_c/\pi$. All plots were generated using the Sellmeier equations given in \cite{Fradkin1999,Konig2004}. We define $\Delta\omega_j=\omega_j-\bar{\omega}_j$ where $\bar{\omega}_j=\omega_p/2$. The number of domains is: a) $N=740$, b) $N=860$, and c) $N=1300$. We note that $N=1300$ is roughly the optimal number of domains needed for a custom-poled crystal, and increasing $N$ does not improve the performance of the algorithm, but rather degrades it. }
\end{center}
\end{figure*}

\section{Results}

In this section, we primarily consider the task of engineering a Gaussian phase-matching function for the purpose of generating pure heralded single photons, and compare our method to existing methods. We also demonstrate that the algorithm can be used to approximate other phase-matching functions of interest.

\subsection{Pure heralded single photons}\label{sec:pure}

Recall that spectral correlations between two down-converted photons necessarily reduce the spectral purity of the individual photons. Here, we show how to dramatically reduce JSA correlations by designing a crystal with a Gaussian phase-matching function. 

For a poled KTP crystal in the type-II configuration, group-velocity matching can be achieved for a crystal with $\Lambda=46\mu m$,  pumped with a $791nm$ laser. In our example, we will work with a pump laser of bandwidth $\sigma=1nm$. 

We first consider a standard periodically poled crystal. For maximal decorrelation, the width of the phase-matching function  should match the width of the pump function, which can be achieved with a periodically poled crystal with $N=740$ domains. Figure 2 a) shows this crystal's phase-matching function, corresponding joint spectral amplitude, and Schmidt coefficients. Notice that the side lobes that arise from the sinc profile of the phase-matching function admit some correlations between the two photons, resulting in a heralded photon purity of $P=0.865$. 

Next, we consider a crystal with a customized duty cycle. Choosing the duty cycle appropriately can drastically reduce the side lobes, as demonstrated by Dixon \emph{et al.} \cite{Dixon2013}, see Figure 2 b). Since modifying the duty cycle broadens the phasematching function, we reduce broadening by increasing the number of domains to $N=860$ to ensure that the widths of the phase-matching and pump functions match. Modifying the duty cycle also shifts the phasematching function, which can be offset by decreasing $\Lambda$, but since this has no effect on the spectral purity, we do not compensate for this. Customizing the crystal's duty cycle increases the heralded photon's purity to $P=0.979$. 
 
Finally, we consider the technique proposed in this paper in which the domain configuration is customized. The technique relies on nontrivial interference between the fields in the crystal to generate a customized phase matching function. In Section \ref{sec:app}, we described how to choose an appropriate $N$ for a given phase matching function. The phase matching function that matches the width of the $1nm$ pump calls for $N=1300$ domains. Note that this $N$ is optimized and increasing the number of domains does not increase the performance of the algorithm, but rather degrades it. In Figure 2 c), we see that for the optimized custom-poled crystal, the side lobes are almost entirely absent and the purity of the heralded photons is increased to $P=0.999$. 

The algorithm optimizes the phase-matching function within the range $[a,b]$, which may result in a non-zero joint-spectral amplitude immediately outside the corresponding frequency range. Since photon-number impurities are only introduced when the filter has non-zero or non-unit transmission in a region of non-zero amplitude, these frequencies can be safely filtered out without compromising the photon-number purity, provided they are sufficiently far from the central peak.

The technique introduced by Bra\'nczyk \emph{et al.} \cite{Branczyk2011}, which  showed that a purity of up to $0.99$ can be achieved by modulating the crystal nonlinearity using higher-order poling, was demonstrated for specific $\sigma$, $\Lambda$, and $N$, and is not readily adaptable to other regimes for direct comparison. 

\subsection{Other functions}

\begin{figure}[t]
\begin{center}
\includegraphics[width=0.95\columnwidth]{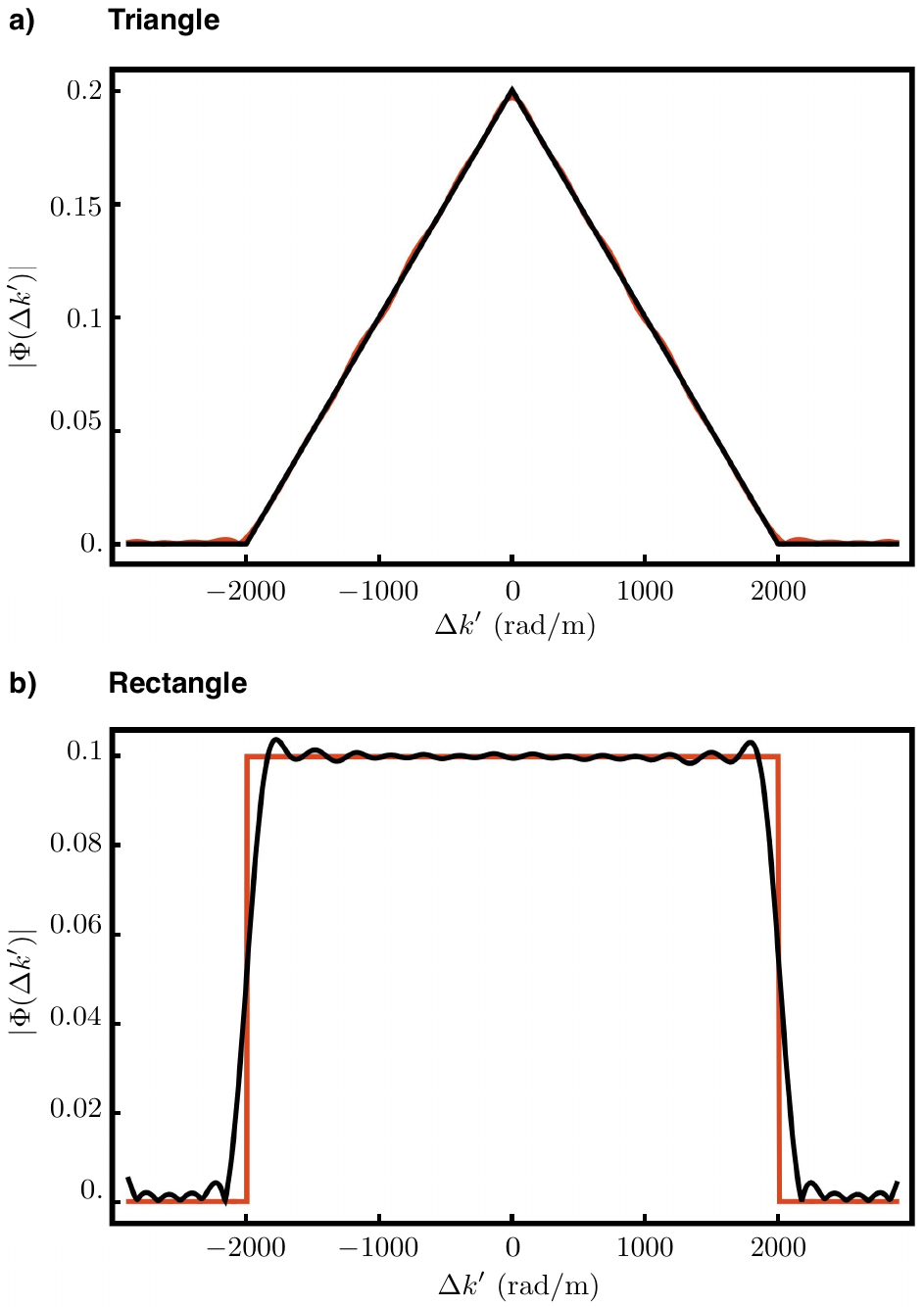}
\caption{(Color online) Phase-matching functions (dimensionless), for: a) triangular and b) rectangular phase-matching functions. Target functions are in red; optimized custom-poled functions are in black. We define the peak nonlinearity of a first-order-poled crystal of the same length as unity and scale all other peak nonlinearities accordingly. We define $\Delta k'=\Delta k-2\pi/\Lambda$. We note that sharp edges make the top-hat function very difficult to approximate with a decomposition of smooth functions. For comparison with other methods, see \cite{Nasr2008,Branczyk2011}.}
\end{center}
\label{fig:3}
\end{figure}

The simulated annealing technique can also be used to approximate other phase-matching functions of interest. Here, we demonstrate this for phase-matching functions with triangular and rectangular profiles. 

Figure 3 a) shows a triangular phase-matching function, generated with an initial configuration of $N=3500$ randomly oriented domains. The algorithm used $4\times10^4$ iterations, with $q=0.999$ and $A=100$, and  \mbox{$M=2001$}  samples to compute the objective function $d_{\bf s}$, over the range $[a,b]=[2\pi(1-0.005)\Lambda,2\pi(1+0.005)\Lambda]$. The height of the target function was $0.4Nl_c/\pi$. 

Due to the sharp edges of a rectangular function, the number of domains needed to produce a reasonable approximation was higher. To generate the phase-matching function shown in Figure 3 b), we increased the number of domains to $N=5000$. We then ran the algorithm with $10^5$ iterations, using $A=1000$. The height of the target function was $0.2Nl_c/\pi$. All other parameters were the same as for the triangle.

\section{Remarks and Future Directions}

We proposed a technique for shaping the phase-matching profile of a pair of downconverted photons by exploiting non trivial interference inside a custom poled crystal. Our method can be used to  approximate profiles of interest, e.g. those with Gaussian, triangular and rectangular profiles. In particular, Gaussian phase matching functions are desirable as they can decorrelate the JSA of downconverted photons.

Because  it eliminates the need for spectral filtering, which reduces the purity of a downconverted squeezed state as a function of the pump power \cite{Branczyk2010,Christ2014}, our technique may facilitate the creation of purer multi-photon states for quantum information processing, e.g. heralded Fock states with high photon number.

Recent work by Quesada and Sipe \cite{Quesada2015}  shows that the joint spectrum for higher Fock states can differ from that of single-photon pairs. It would  be interesting to apply our technique in this regime. It might also be possible to apply our technique to the optimization of the entire joint spectral amplitude, rather than just the phase-matching function, as was shown by Phillips \emph{et al.} \cite{Phillips2013}. While simulated annealing was sufficient for solving the present problem, more sophisticated algorithms, such as genetic algorithms, might be necessary for further extensions. 

We also expect our technique to have applications in classical nonlinear optics, such as second harmonic generation, where similar spectral shaping techniques have been demonstrated \cite{Fejer1992,Imeshev2001}.

\section{Acknowledgements}
AMB thanks Alessandro Fedrizzi, Michele Mosca, Thomas Stace and Phillip Vetter for interesting discussions. We thank Ben Dixon for providing thoughtful comments on our manuscript. We also thank Marc Burns for suggesting ways to optimize our code. AD acknowledges financial support from the Institute for Quantum Computing undergraduate research program. This research was supported in part by Perimeter Institute for Theoretical Physics. Research at Perimeter Institute is supported by the Government of Canada through Industry Canada and by the Province of Ontario through the Ministry of Research and Innovation. LC acknowledges support from the John Templeton Foundation. 

\appendix
\section{C++ code}

A \verb+C+++ implementation of the algorithm described in this paper is available in the ancillary files to this submission. The folder contains a `\verb+readme.rtf+' file which details how to compile and run the code. The zipped package can also be downloaded from \url{www.agatabranczyk.com/Dosseva2015code.zip}.

\end{document}